\documentclass[aps,prr,superscriptaddress,reprint,amsmath,amssymb,aps,longbibliography]{revtex4-2}
\usepackage{graphicx}
\usepackage{float}
\usepackage[
pdfstartview=FitH,
bookmarks=true,
breaklinks=true,
colorlinks=true,
linkcolor=blue,anchorcolor=blue,
citecolor=blue,filecolor=blue,
menucolor=blue,urlcolor=blue]{hyperref}

\usepackage{xcolor}
\usepackage{ulem}

\begin{document}

\title{Strong interaction induced dimensional crossover in 1D quantum gas}

\author{Zhongchi Zhang}
\thanks{These authors contribute equally to this work.}
\author{Zihan Zhao}
\thanks{These authors contribute equally to this work.}
\author{Huaichuan Wang}
\author{Ken Deng}
\author{Yuqi Liu}
\affiliation{Department of Physics and State Key Laboratory of Low Dimensional Quantum Physics, Tsinghua University, Beijing, 100084, China}
\author{Wenlan Chen}
\email{cwlaser@ultracold.cn}
\affiliation{Department of Physics and State Key Laboratory of Low Dimensional Quantum Physics, Tsinghua University, Beijing, 100084, China}
\affiliation{Frontier Science Center for Quantum Information and Collaborative Innovation Center of Quantum Matter, Beijing, 100084, China}
\author{Jiazhong Hu}
\email{hujiazhong01@ultracold.cn}
\affiliation{Department of Physics and State Key Laboratory of Low Dimensional Quantum Physics, Tsinghua University, Beijing, 100084, China}
\affiliation{Beijing Academy of Quantum Information Science, Beijing, 100193, China}
\affiliation{Frontier Science Center for Quantum Information and Collaborative Innovation Center of Quantum Matter, Beijing, 100084, China}

\begin{abstract}
 We generated a one-dimensional quantum gas confined in an elongated optical dipole trap instead of 2D optical lattices. The sample, comprising thousands of atoms, spans several hundred micrometers and allows for independent control of temperature and chemical potential  using Feshbach resonance. 
This allows us to directly observe and investigate the spatial distribution and associated excitation of 1D quantum gas without any ensemble averaging. In this system, we observed that the dimension of 1D gas will be popped up into 3D due to strong interaction without changing any trapping confinement. During the dimensional crossover, we found that increasing the scattering length leads to the failure of 1D theories, including 1D mean field, Yang-Yang equation, and 1D hydrodynamics.
Specifically, the modified Yang-Yang equation effectively describes this 1D system at temperatures beyond the 1D threshold, but it does not account for the effects of stronger interactions.
Meanwhile, we observe two possible quantized plateaus of breathing-mode oscillation frequencies predicted by 1D and 3D hydrodynamics, corresponding to weak and strong interactions respectively. And there is also a universal crossover connecting two different regimes where both hydrodynamics fail.
\end{abstract}
\maketitle

\section{Introduction}
One-dimensional (1D) physics displays numerous of unique fascinating phenomena and attracts plenties of research due to its unique 1D integrability \cite{GiamarchiBook2003,2017LNP,kinoshita_quantum_2006,tang_thermalization_2018,PhysRevLett.122.090601}. 
Among these, the ultracold-atom system is one of the most important platforms to demonstrate and investigate these low-dimensional physics \cite{RevModPhys.83.1405,Guan_2022}. Under the condition of 1D, varieties of new physics are demonstrated such as bosonic fermionization \cite{PhysRevA.66.053614,PhysRevA.66.053613,PhysRevLett.86.5413,Bel2004Tonks,science.aaz0242}, meta-stable highly excited state \cite{Super_TG_Gas_science,PhysRevLett.95.190407,PhysRevLett.105.175301,PhysRevA.81.031609,Science_Benjamin_lev}, non-equilibrium dynamics \cite{le_observation_2023,PhysRevLett.119.165701,zhao2023observation}, spin-charge separations \cite{Bloch_science,PhysRevLett.125.190401,science.abn1719}, and solitons \cite{Bright_soliton_science_2002,2002Formation,PhysRevLett.89.200404,PhysRevLett.83.5198,Dark_soliton_science_2000,nguyen_collisions_2014,science.aal3220}.
Among these effects, the dimension plays a central role. Therefore, it is important to investigate and understand how the dimension influences a quantum system and how non-trivial phenomena emerge when a system at higher dimension is reduced into 1D \cite{PhysRevLett.92.130403,PhysRevLett.105.265302,PhysRevA.83.021605,PhysRevLett.106.230405,PhysRevLett.130.123401}.

\begin{figure*}[tb]
\centering
\includegraphics[width=0.95\textwidth]{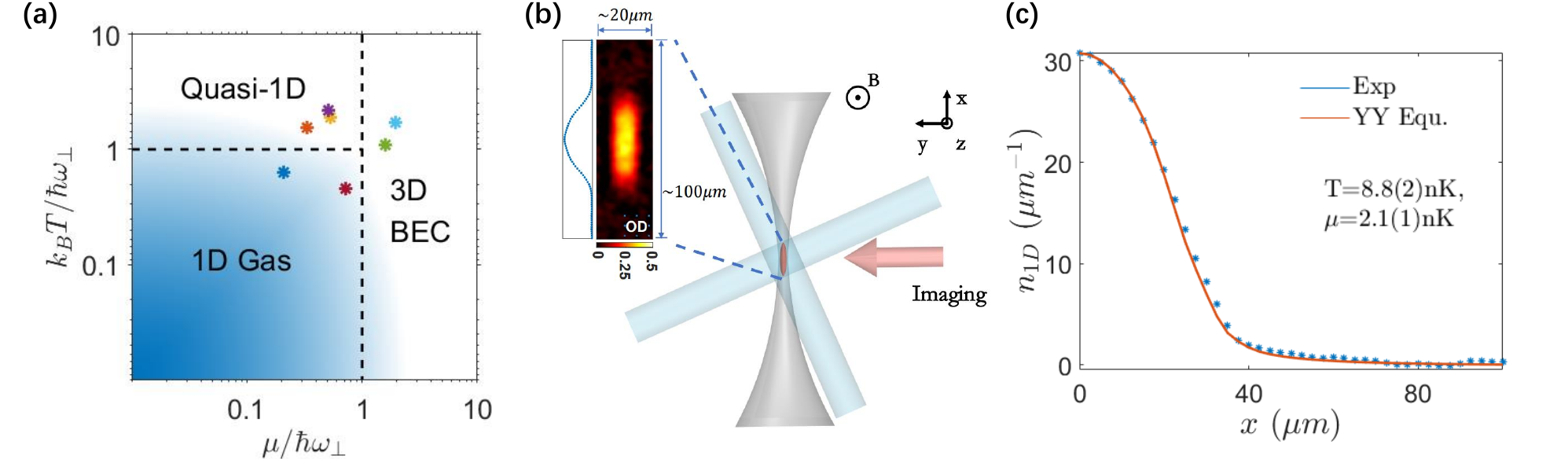}
\caption{(a) Phase diagram of atoms trapped in an elongated trap. The system's behavior is governed by three pivotal parameters: the radial confinement's vibrational frequency $\omega_{\perp}$, the temperature $T$, and the chemical potential $\mu$. In conditions where $k_b T, \mu \ll \hbar \omega_{\perp}$, the system adopts a 1D regime, described by the Yang-Yang equation \cite{10.1063/1.1664947}. In contrast, for $\mu \ll \hbar \omega_{\perp}$ and $k_b T \gtrsim \hbar \omega_{\perp}$, the system manifests as a mixed state comprising multiple 1D gases, each populating a unique harmonic level within the radial confinement, thus forming a quasi-1D gas ensemble.  
Moreover, when $\mu \gtrsim \hbar \omega_{\perp}$, even at zero temperature, the 1D approximation becomes invalid, transitioning the system into a 3D BEC regime.The points on the graph correspond one-to-one with the lines in Fig.~\ref{Fig3}, with matching data represented by the same color. (b) The experimental setup. A highly-focused running wave laser at 1560~nm (gray) along the $x$-axis yields tight confinement in the $y$- and $z$-directions. Then a weak crossed dipole trap at 1064~nm provides the tunability of $\omega_\parallel$. The imaging direction is along the $y$-axis. A typical imaging data is shown, where we see a significant expansion along the $z$-axis due to 3~ms flight, with negligible change in the density distribution along the $x$-axis. The following analysis centers on the integrated 1D density distribution along the $x$-axis. 
(c) A typical density distribution of 1D-gas measured in our experiment with $\omega_{\parallel}=2\pi\times$3.0(1)~Hz and $a_s=60a_0$, showing our ability to generate real-1D gas.
}
\label{Fig1}
\end{figure*}

\begin{figure}[hbtp] 
\centering
\includegraphics[width=0.4\textwidth]{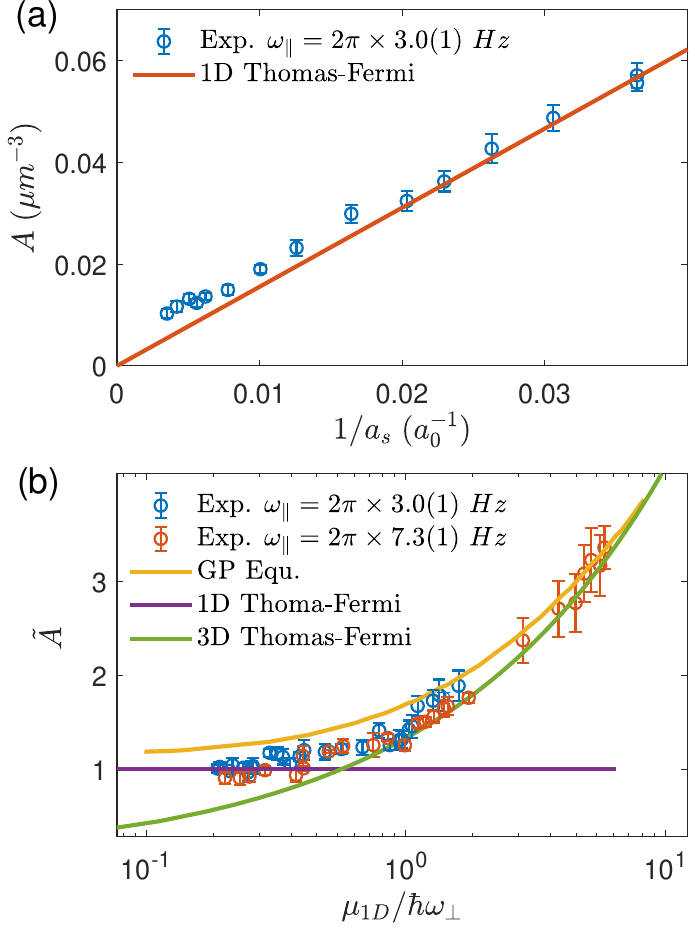}
\caption{
(a) The quadratic coefficient $A$ is plotted against $1/a_s$ in a trap with $\omega_\parallel=2\pi \times3.0(1)$~Hz. 
The atom number $N$ ranges from $1.3 \times 10^3$ to $7.7\times 10^3$ due to the three-body loss during the preparation. 
The solid red line is the theoretical result based on the 1D Thomas-Fermi distribution and the deviation indicates a departure of the system from ideal 1D behavior.
(b) The dimensionless coefficient $\tilde{A}$ versus the effective 1D chemical potential $\mu_{1D}$. Here the horizontal axis is labeled by $\mu_{1D}/\hbar\omega_\perp$ which is also dimensionless. 
Data represented by different colors, corresponding to varying $\omega_{\parallel}$ values (3~Hz for blue and 7.3~Hz for red), show similar behaviors around $\mu_{1D}\approx\hbar\omega_\perp$.
Solid yellow, purple, and green lines correspond to theoretical calculations using the 3D Gross-Pitaevskii equation, 1D Thomas-Fermi distribution, and 3D Thomas-Fermi distribution, respectively.
}
\label{Fig2}
\end{figure}

Therefore, it is intuitive for people to study the dimensional crossover by tailoring the confinement, such as slicing 3D bulk materials into 2D films or 1D lines, and during this procedure, new phenomena emerge \cite{PhysRevLett.130.123401,PhysRevA.70.013606,PhysRevLett.87.130402,PhysRevLett.113.215301,PhysRevResearch.5.013136}.
However, the modification of geometry represents only a portion of the underlying phenomena.
In fact, the interaction between particles is also effecting the dimension of a quantum system \cite{PhysRevA.66.053614}. 
Here, we demonstrate that even without changing the confinement geometry, the quantum gas can still display different dimensional properties in 1D or 3D where strong interaction directly pops up a 1D quantum gas into a 3D BEC. Specifically, this transition to a higher-dimensional state is universally driven by interactions, distinct from thermal excitations caused by temperature changes. This assertion is experimentally supported by two evidences. 
The first evidence, derived from measuring spatial distribution, shows that the modified Yang-Yang equation \cite{van_amerongen_yang-yang_2008}, adept at describing 1D physics, is applicable in both low and high temperature regimes but fails in the strong-interaction regime.
The second evidence is derived from the measurement of breathing-mode frequencies at different interaction strengths. Here, we observe two possible plateaus, aligned with 1D and 3D hydrodynamic predictions \cite{PhysRevA.66.043610} for respective interaction strengths. Notably, a universal crossover regime exists where both hydrodynamic models fail.

One improvement is that we generate one piece of 1D quantum gas in an elongated optical dipole trap, where temperature and chemical potential can be both smaller than the vibrational frequency of tightly-confined directions. Comparing to the atom-chip experiments utilizing a magnetic trap, we can tune temperatures, chemical potentials, and scattering lengths independently in the optical trap by utilizing a magnetic Feshbach resonance \cite{RevModPhys.82.1225}. 
Comparing to preparing 1D gas in a 2D optical lattice \cite{Bel2004Tonks,science.aaz0242,Super_TG_Gas_science,Science_Benjamin_lev,le_observation_2023,PhysRevLett.119.165701,science.abn1719,science.abf0147,meinert_probing_2015,PhysRevLett.92.130405,zhao2023observation, Bloch_science,PhysRevLett.125.190401}, 
our 1D sample contains thousands of atoms, and it's suitable for direct in-situ measurements to retrieve the spatial information. 
Usually the measurement of 1D gas in optical lattices is probing an ensemble averaging of thousands of samples at one time where atoms are inhomogeneously distributed. 
By utilizing our method, we can probe 1D gas directly without any ensemble averaging.

\section{Experimental Setup}
We prepare rubidium-85 BEC in the hyperfine level $|F=2,m_F=-2\rangle$ by the gravity-forced evaporative cooling \cite{KETTERLE1996181,Saptarishi_Chaudhuri_2007}. 
After that, we transfer atoms to an elongated dipole trap consisted of a running wave and a crossed dipole trap.
The running wave has a wavelength 1560~nm and a beam waist 25 $\mu$m. At a power of 37 mW, it provides a radial vibrational frequency $\omega_{\perp}/2\pi$=230~Hz =$k_B/h\times$ 11.5 ~nK in two tightly-confined directions ($y$ and $z$).
The crossed dipole trap is formed by two weak beams at a wavelength of 1064~nm with a beam waist 100 $\mu$m, allowing the tuning of the longitudinal vibrational frequency $\omega_{\parallel}/2\pi$ along the $x$-axis within a range of 3~Hz to 17~Hz. 
During this stage, we adjust the scattering length to a desired value $a_s$ for the following experiments based on the Feshbach resonance. By tuning the evaporation parameters, we can control the final atom number $N$ in a range of $10^3$ to $10^4$ and temperatures within a range of 7 to 100~nK.

First, we would like to prove the ability of creating real-1D gas in the elongated dipole trap. For each sample, we measure the spatial density distribution by absorption imaging \cite{PhysRevApplied.20.014037} and obtain the 1D density profile along $x$ axis. 
Here we show data of 1D density $n_{1D}$ with $\omega_{\parallel}/2\pi=3$~Hz and $a_s=60a_0$ in Fig. \ref{Fig1}(c). 
By utilizing the Yang-Yang equation \cite{10.1063/1.1664947} and local density approximation (LDA) $n_{1D}(x)=n_{YY}(\mu(x),T)$ \cite{PhysRevA.58.1563,PhysRevA.80.043605,PhysRevB.94.205134}, we obtain the temperature $T=$8.8(2)~nK and the chemical potential $\mu=2.1(1)$~nK, which confirms that both temperature and chemical potential are less than $\hbar\omega_{\perp}/k_B=11.5$~nK. Meanwhile, the density distribution is fully consistent with the Yang-Yang equation which precisely describes the 1D physics. The dimensionless interaction parameter $\gamma=mg_{1D}/(\hbar^2n_0)\approx 4\times10^{-4} \ll1$ indicates that the central part of the system is in the weakly interacting regime,  
where $g_{1D}=2\hbar\omega_{\perp}a_s$ is the effective 1D interaction strength, and $n_0$ is the central density of atoms.

\begin{figure*}[hbtp]
\centering
\includegraphics[width=0.9\textwidth]{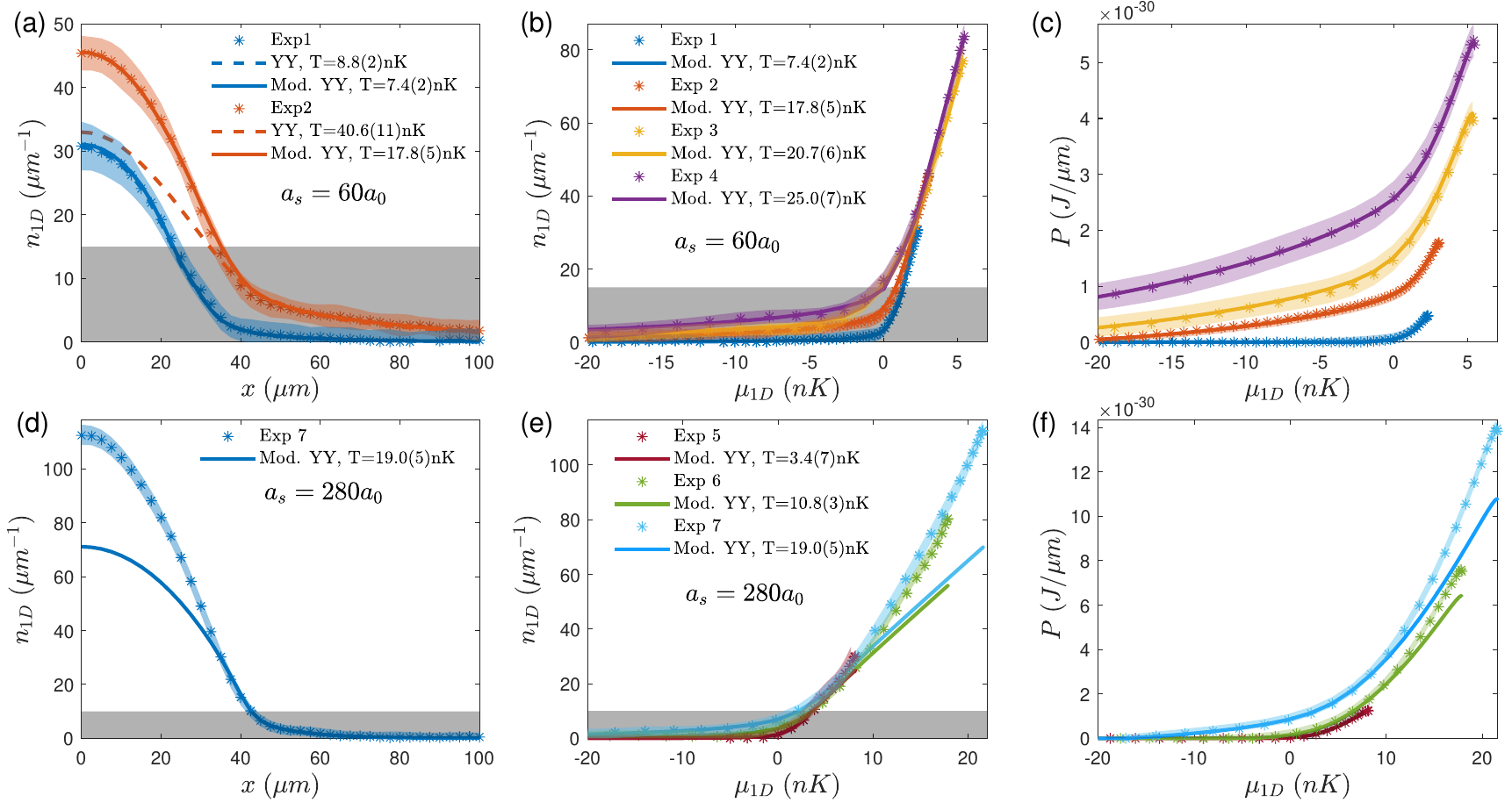}
\caption{Benchmarking modified Yang-Yang equation in different regimes with $\omega_\perp=2\pi\times$230~Hz. (a), (b), and (c) correspond to atom in the 1D and quasi-1D regimes with $a_s=60a_0$. 
(d), (e), and (f) correspond to atom in the 3D BEC regime with $a_s=280a_0$.
The gray shaded area represents the section used for fitting. The colored shaded area represents the $\pm 1 \sigma$ error bars of the experimental data, encompassing both the statistical fitting error and the uncertainty in $\omega_{\parallel}$.
(a) and (d) are the spatial density distributions. (b) and (e) are EoS of 1D density $n_{1D}$ versus the chemical potential $\mu_{1D}$ (see Appendix~\ref{Appendix:EOS}). The region where the chemical potential is less than zero is mainly occupied by thermal atoms.
(c) and (f) are EoS of pressure $P$ versus $\mu_{1D}$ where $P(\mu,T)=\int_{-\infty}^\mu n(\mu^\prime,T)d\mu^\prime$ \cite{PhysRevResearch.5.013136}. In each plot, scattered points are measured data and lines are fitted curves based on either Yang-Yang equation (dashed) or modified one (solid).
}
\label{Fig3}
\end{figure*}

\section{1D to 3D crossover}
Once we confirm the creation of 1D quantum gas at low temperatures, we want to investigate how 1D gas becomes a 3D BEC due to interaction. 
For atoms with sufficiently low temperatures, the central spatial distribution is described by the Thomas-Fermi distributions either in 1D or 3D regime depending on the interaction strength \cite{PhysRevA.66.043610}. For small interaction, the 1D Thomas-Fermi distribution gives a density distribution as $n(x)=(\mu_0-m\omega^2_\parallel x^2/2)/g_{1D}$. The coefficient of the quadratic term $x^2$ is inversely proportional to $a_s$ and independent on the chemical potential $\mu_0$. On the other hand, the integrated 3D Thomas-Fermi distribution predicts $n(x)=\pi (\mu_0-m\omega^2_\parallel x^2/2)^2/(m\omega^2_\perp g_{3D})$, where $g_{3D}=4\pi\hbar^2a_s/m$ is the 3D interaction strength (see Appendix~\ref{Appendix:DensityDistributions}). It is not a quadratic distribution, but if we fit it with a quadratic trial function, the coefficient of the fitted quadratic term will strongly depend on the chemical potential.
Therefore, we change the scattering length $a_s$ from $28 a_0$ to $280a_0$ and measure the 1D density distribution with $\omega_\parallel=2\pi\times3.0$~Hz or $2\pi\times7.3$~Hz. The three-body loss limits the ability of achieving a very large scattering length (see Appendix~\ref{Appendix:3BodyLoss}). Then we fit 1D density with a quadratic function $n(x)=-A x^2+B$. Here we only use the central one-third atoms for this fitting to exclude the influence of thermal parts \cite{PhysRevLett.119.165701,PhysRevA.62.023604}. The corresponding results are shown in Fig. \ref{Fig2}(a). The deviation of data from the trend predicted by the 1D Thomas-Fermi distribution indicates a departure of the system from an ideal 1D behavior. 

For a better illustration, we utilize a dimensionless parameter $\tilde{A}=A\times4\hbar\omega_\perp a_s/m\omega^2_\parallel$. 
As shown in Fig.~\ref{Fig2}(b), $\tilde{A}=1$~when the system is in the 1D regime ($\mu_{1D}<\hbar\omega_{\perp}$). Here $\mu_{1D}$ is the effective 1D chemical potential $\mu_{1D}=g_{1D}n_0$. 
Then $\tilde{A}$ deviates from 1 when $\mu_{1D} \gtrsim\hbar\omega_{\perp}$.
Here we list two sets of data with different $\omega_\parallel$ ($2\pi\times$3~Hz and $2\pi\times$7.3~Hz). 
Two sets of data show similar behaviors that the crossover from 1D to 3D happens near $\mu_{1D}/\hbar\omega_\perp\sim 1$.
Meanwhile, we also plot the lines predicted by the 1D Thomas-Fermi distribution, the 3D one, and the numerical results from three-dimensional Gross-Pitaevskii equation (3D GPE) \cite{PhysRevE.62.1382,BAO2003318,ANTOINE20132621}. 
This measurement shows strong interaction will change the shape of density distribution and pop up the dimension into 3D.

After measuring the central part dominated by interactions, we conduct a more detailed analysis by considering temperature effects and give quantitative differences between 1D gas, quasi-1D gas, and 3D BEC. In Fig.~\ref{Fig3}(a) (blue solid line), we show the measured density distribution in 1D regime where the Yang-Yang equation matches data.
When $k_B T\gtrsim \hbar\omega_\perp$, we have to include the thermal contributions from the excited levels along $y$ and $z$ directions. Therefore, we adopt the modified Yang-Yang equation \cite{van_amerongen_yang-yang_2008} to describe these effects.
Here we consider atoms as two parts. The primary part comprises atoms in the radial ground state described by the Yang-Yang equation. The remaining component consists of atoms in the radial excited states with each state treated as an ideal 1D Bose gas that only interacts with atoms in the radial ground state. Then, using LDA, we incorporate the longitudinal potential by introducing a varying chemical potential $\mu(x)=\mu_0-V(x)$. 
Within this model, the linear density is given by:
\begin{equation}
    n_{1D}[\mu(x),T]=n_\mathrm{YY}[\mu(x),T]+\sum_{j=1}^\infty(j+1)n_{j}[\mu_j(x),T],
\end{equation}
where $n_j(\mu_j,T)=g_{1/2}[\exp(\mu_j/k_BT)]/\Lambda_T$  represents the 1D density of radial excited state with quantum number $j\geq1$ and degeneracy $j+1$. $g_{1/2}$ is a polylog function accounting for Bose integration, $\Lambda_T=(2\pi\hbar^2/mk_BT)^{1/2}$ is the thermal de Broglie wavelength, and $\mu_j(x)=\mu(x)-j\hbar\omega_\perp -g_{1d}n_{YY}(x)$ is the local chemical potential for each radial excited state.

To benchmark the consistency of Yang-Yang equation and modified one, we only use the low density part ($\mu(x)=g_{1D}n(x)<3$~nK indicated by shadow areas in Fig.~\ref{Fig3}) to determine the central chemical potential $\mu_0$ and temperature $T$. We then compare the measured density distribution with the predictions of both the Yang-Yang equation and the modified Yang-Yang equation.
For 1D gas (Fig.~\ref{Fig3}(a), blue line), both of the Yang-Yang equation and the modified Yang-Yang equation give consistent results. The temperatures are slightly different due to model dependence. 
For quasi-1D gas where $k_B T > \hbar\omega_{\perp}$ but $\mu_0 < \hbar\omega_{\perp}$, the modified Yang-Yang equation still yields consistent results with experimental data, including high-density distributions and the equation of state (EoS), while the Yang-Yang equation fails due to its lack of consideration for the thermal wing (Fig.~\ref{Fig3}(a) to (c)).
For 3D BEC, the Yang-Yang equation and modified one both fails to describe the high density parts of atoms (Fig.~\ref{Fig3}(d) to (f)). Particularly,
the deviation from the modified Yang-Yang equation becomes evident around $\mu_0 \simeq \hbar\omega_\perp=k_B\times 11.5$~nK, indicating that the interaction energy promotes atoms to higher radial vibrational states and this mechanism is different with the thermal excitation. It is noteworthy that the precise location of this deviation appears to be temperature-independent within our system.

The examination of low-energy excitation is also one significant approach for investigating the properties of complex systems. 
Here, we excite the breathing mode by quenching the scattering length $a_s$ from $a_s^0$ to $a_s^D$. To maintain a consistent breathing mode amplitude below 20\% \cite{PhysRevLett.77.988,PhysRevLett.77.420,PhysRevLett.113.035301}, we keep the quench strength $\alpha\equiv a_s^0/a_s^D$ at around 2.
This amplitude is sufficiently small to ensure we are in the perturbation regime. 
We perform the in-situ images after an evolution time $\tau$. 
By fitting the $x$-direction atomic waist versus time $\tau$, we obtain the oscillation frequency $\omega_B$ of the breathing mode (see Appendix~\ref{Appendix:BreathingModeFreq}).
When atoms are in the strong interaction regime and described by an elongated 3D BEC, 
the hydrodynamics gives a prediction of $\omega_B/\omega_{\parallel} =\sqrt{5/2}\approx1.581$  \cite{PhysRevLett.77.2360,PhysRevA.66.043610,PhysRevLett.77.988}. Conversely, in the 1D regime, where $\hbar \omega_\parallel \ll \mu \ll \hbar \omega_\perp$, it results  $\omega_B/\omega_\parallel=\sqrt{3}\approx 1.73$  \cite{PhysRevA.58.2385,PhysRevA.59.1477,1999Quasi,PhysRevA.66.043610,PhysRevLett.113.035301}. As we continue to reduce the interaction, the system enters the ideal Bose gas regime where $\mu < \hbar \omega_\parallel$. In this regime, a single-particle description is sufficient and $\omega_B=2\omega_\parallel$. 

\begin{figure}[hbtp]
\centering
\includegraphics[width=0.42\textwidth]{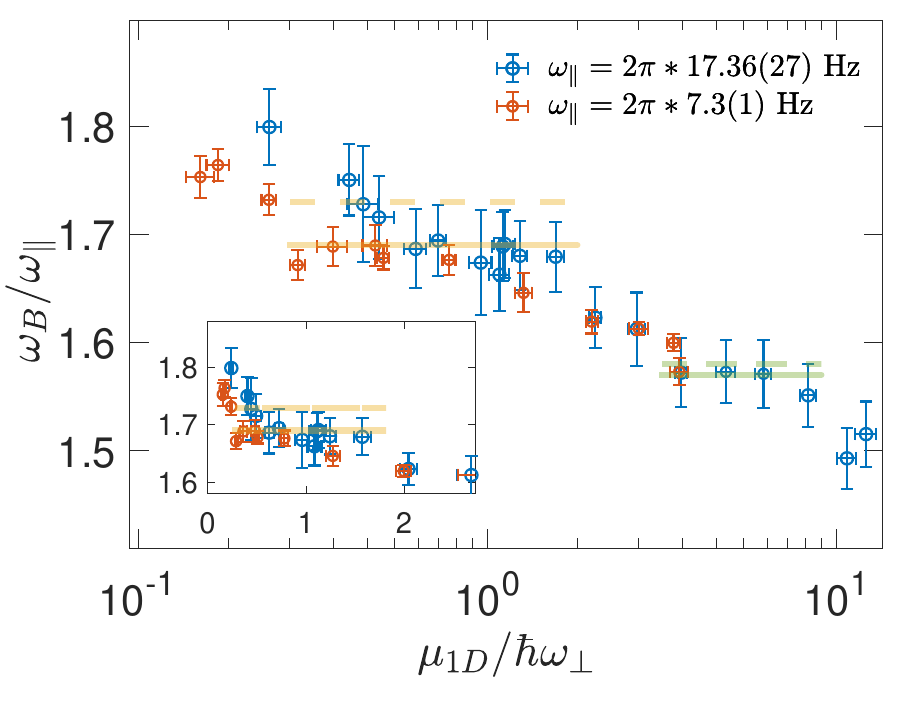}
\caption{Breathing-mode frequencies through the crossover in a logarithmic (main panel) and linear (inset) plot. Two data sets are measured for different axial frequencies, $\omega_{\parallel} = 2\pi \times 17.36$ Hz (blue circles) and $2\pi \times 7.3$ Hz (red circles), respectively, but with the same transverse frequency, $\omega_{\perp} = 2\pi \times 230$ Hz. The atom number is $N = 2.4(7) \times 10^3$ and the scattering length $a_s$ ranges from $7.5a_0$ to $400a_0$.
By rescaling both axes to dimensionless units,
different sets of data converge into one curve for $\mu_{1D}/\hbar\omega_\perp\ge0.5$.
Two possible quantized plateaus emerge in the 1D (yellow solid line with a value 1.69) and 3D (green solid line with a value 1.57) regimes. 
The dashed lines corresponds to
the theoretical predicted values based on perturbation theories in 1D (yellow dashed line with $\sqrt{3}$) and 3D (green dashed line with $\sqrt{5/2}$) hydrodynamics.
}
\label{Fig4}
\end{figure}

In Fig.\ref{Fig4}, we present the data of $\omega_B/\omega_\parallel$ versus $\mu_{1D}/\hbar\omega_\perp$ with two different $\omega_\parallel$ (7.3~Hz and 17.4~Hz). By tuning scattering lengths $a_s^D$ (7.5 to 400$a_0$), we change the interaction of atoms and cross different phase regimes. 
The ratio $\omega_B/\omega_\parallel$ exhibits two possible plateaus (1.57 and 1.69, solid lines in Fig.~\ref{Fig4}) at $\mu_{1D}/\hbar\omega_{\perp}>4$ and $\mu_{1D}/\hbar\omega_{\perp}<1$, corresponding to the 3D and 1D regimes respectively.
Small discrepancies (0.6\% and 2.3\%) between the measured values and predicted one have also been reported in the previous experiments \cite{PhysRevLett.77.988,PhysRevLett.77.420,PhysRevLett.113.035301}. 
For very small interaction where $\mu_{1D}$ is comparable with the longitudinal vibrational frequency $\hbar \omega_{\parallel}$, the frequency $\omega_B$ of the breathing modes further increases because the atoms are more like ideal Bose gas. Despite our experimental parameters not achieving the ideal-Bose-gas limit, we still observe effects attributable to this mechanism. Comparing two data sets, the left turning points of the 1D plateau are different. In systems with larger $\omega_{\parallel}$ (blue data in Fig. \ref{Fig4}), as the chemical potential decreases, the criterion $\mu_{1D}/\hbar\omega_{\parallel} \sim 1$ is satisfied earlier, leading to an increase in $\omega_B/\omega_\parallel$ at a higher $\mu_{1D}$. This shows transitions approaching to ideal Bose gas.
The right turning points of the 1D plateau have the same position because the crossover of 1D to 3D happens at $\mu_{1D}/\hbar\omega_{\perp}\sim 1$, independent of $\omega_\parallel$. Meanwhile, there is also a decreasing of $\omega_B$ at $\mu_{1D}/\hbar\omega_{\perp}\approx 8$. The observed decrease is attributed to the significant damping of oscillations and three-body loss (see Appendix \ref{Appendix:3BodyLoss}), a consequence of three-body recombination processes facilitated by a large scattering length \cite{PhysRevLett.77.2921,PhysRevLett.83.1751,PhysRevLett.91.123201,PhysRevA.84.033632}. And the crossover from 1D to 3D hydrodynamics is universal where two sets of data match each other when $\mu_{1D}/\hbar\omega_\perp$ ranges from 0.5 to 5. 

\section{Conclusions}
In conclusions, we create 1D quantum gas in an elongated dipole trap with independently-tuned temperatures and chemical potential, and we demonstrate the dimensional crossover from 1D to 3D due to the strong interaction where 1D theories gradually fail. 
Our results suggest that the dimension of a physical system cannot be simply defined by geometry or confinements and there is also a crossover regime where the interaction strongly dominates physical properties including ground states and low-energy excitation. 
We hope this work can inspire further investigations in strongly-correlated low-dimensional systems, potentially where there are emerging phenomena cannot be classified into any integer-defined dimensional properties.

\begin{acknowledgments}
This work is supported by National Key Research and Development Program of China (2021YFA0718303, 2021YFA1400904) and National Natural Science Foundation of China (92165203, 61975092, 11974202).
\end{acknowledgments}

\appendix

\section{Density distributions}
\label{Appendix:DensityDistributions}
\subsection{One-dimensional regime}
For 1D gas, where the chemical potential and temperature satisfy $\hbar\omega_\parallel\ll\mu_{1D}\ll\hbar\omega_\perp$, $k_B T\ll\hbar\omega_\perp$,  the thermodynamic behavior can be fully described by the Yang-Yang equation \cite{10.1063/1.1664947}.

\begin{equation}
\varepsilon(k)=\frac{\hbar^2k^2}{2m}-\mu-\frac{k_BTc}{\pi}\int_{-\infty}^{\infty}dq\frac{\ln\left(1+\mathrm{e}^{-\frac{\varepsilon(q)}{k_BT}}\right)}{c^2+(k-q)^2},
\end{equation}
where $k$ is the quasi-momentum defined by the Bethe Ansatz wavefunction and $\epsilon_k$ is the dressed energy. $c={mg_{1D}}/{\hbar^2}$, with $g_{1D}=2\hbar\omega_\perp a_s/(1-1.46a_s/a_\perp)$. Here, $a_\perp=\sqrt{2\hbar/m\omega_\perp}$ is the characteristic length of radial harmonic trap and $a_s$ is the three-dimensional scattering length. Given a chemical potential $\mu$ and temperature $T$, we can numerically solve this iterative equation and get the dressed energy $\epsilon(k)$. Then we can get the pressure via 
\begin{equation}
p=k_{B}T\int_{-\infty}^{\infty}dk\ln(1+\mathrm{e}^{-\varepsilon(k)/k_{B}T})/2\pi
\end{equation}
and the 1D density via $n(\mu)=\partial p/\partial\mu$. By using the local density approximation (LDA) $\mu(x)=\mu_0-V(x)$, we can get the whole 1D density distribution based on the spatial coordinate $x$. 

When the temperature gets higher, the population of radial excited levels becomes important. By treating these atoms as non-interacting boson, the system is described by the modified YY equation \cite{van_amerongen_yang-yang_2008}.
\begin{equation}
    n_{1D}[\mu(x),T]=n_\mathrm{YY}[\mu(x),T]+\sum_{j=1}^\infty(j+1)n_{j}[\mu_j(x),T],
\end{equation}
where $n_j(\mu_j,T)=g_{1/2}[\exp(\mu_j/k_BT)]/\Lambda_T$  represents the 1D density of radial excited state with quantum number $j\geq1$ and degeneracy $j+1$. $g_{1/2}$ is a polylog function accounting for Bose integration, $\Lambda_T=(2\pi\hbar^2/mk_BT)^{1/2}$ is the thermal de Broglie wavelength, and $\mu_j(x)=\mu(x)-j\hbar\omega_\perp -g_{1d}n_{YY}(x)$ is the local chemical potential for each radial excited state. 

When the dimensionless interaction parameter $\gamma=\epsilon_{int}/\epsilon_{kin}=mg_{1D}/\hbar^2 n_{1D}$ is much smaller than 1, the system enters the 1D mean-field regime \cite{PhysRevA.66.043610}. In this regime, the 1D GP equation is sufficient for describing the density distribution at zero-temperature limit, where the 1D GP equation is written as
\begin{equation}
i \hbar \frac{\partial \psi(x,t)}{\partial t} = \left( -\frac{\hbar^2}{2m} \frac{\partial^2}{\partial x^2} + V(x) + g_{1D} |\psi(x,t)|^2 \right) \psi(x,t).
\end{equation}
By ignoring the effect of temperature and kinetic energy, we can use the 1D Thomas-Fermi distribution to capture the main feature of this system,
\begin{equation}
    n(x)=\frac{\mu_0-m\omega_\parallel^2x^2/2}{g_{1D}}\approx -\frac{m\omega_\parallel^2}{4\hbar\omega_\perp a_s}x^2+\frac{\mu_0}{2\hbar\omega_\perp a_s},
\end{equation}
where we ignore the confinement induced resonance (CIR) term due to $a_s\ll a_\perp$. Therefore, the observable that used in Fig. 2, the coefficient of the quadratic term $x^2$ is $A={m\omega_\parallel^2}/\left({4\hbar\omega_\perp a_s}\right)$, which is inversely proportional to $a_s$ and independently on either the chemical potential $\mu_0$ or atom number $N$. Furthermore, if we utilize a dimensionless parameter $\tilde{A}=A\times4\hbar\omega_\perp a_s/m\omega^2_\parallel$, the coefficients of all 1D data should be equal to 1 when the system is in the 1D regime, regardless of the values of $\omega_\parallel$, $\omega_\perp$, and $a_s$.

\subsection{Three-dimensional regime}
For strong interaction strength, where the chemical potential satisfies $\mu_{1D}\gg\hbar\omega_\perp$, the density distribution in a 3D harmonic trap should obey the 3D Thomas-Fermi distribution in the zero-temperature limit, and the density distribution $n(x,y,z)$ is written as
\begin{equation}
    n(x,y,z)=\frac{\mu_0-m(\omega_\parallel^2x^2+\omega_\perp^2y^2+\omega_\perp^2z^2)/2}{g_{3D}},
\end{equation}
where $g_{3D}=4\pi\hbar^2a_s/m$. Then we integrate the three-dimensional density over the radial directions to obtain the one-dimensional density distribution along the axial direction, expressed as $n(x)=\iint n(x,y,z)\,dy\,dz$, which gives,
\begin{equation}
    n(x)=\pi (\mu_0-m\omega^2_\parallel x^2/2)^2/(m\omega^2_\perp g_{3D}).
\end{equation}
The given polynomial represents a quartic function. Thus, attempting to fit it using a quadratic function leads to the inability to express the coefficient of the quadratic term analytically. Consequently, we are limited to a numerical comparison between the experimental data and the theoretical predictions as depicted in Fig. 2(b). 

\subsection{1D to 3D crossover regime}
Unfortunately, for intermediate interaction strength, where the chemical potential satisfies $\mu_{1D}\sim\hbar\omega_\perp$, there is no analytic expression for density distribution, which is one of the reasons why people are interested in this regime. The only theoretical tool we can use is the 3D GPE, which predicts that all the data in Fig. 2(b) with different $\omega_\parallel$ will collapse into a single curve in this regime.


\section{Equation of state of the system}
\label{Appendix:EOS}
To compare the equation of state (EOS) from experimental data with predictions of the modified Yang-Yang equation, we focus on the low-density regime ($\mu(x) = g_{1D} n(x) < 3$ nK), as indicated by shaded areas in Fig. 3. This region is used to determine the central chemical potential $\mu_0$ and temperature $T$. By minimizing the residual sum of squares ($RSS = \sum e^2_i = \sum \left[n_{\text{exp}}(x_i) - n_{\text{Mod.YY}}(\mu_0, T, x_i)\right]^2$), we obtain optimal-estimated $\tilde{\mu_0}$ and $\tilde{T}$. The confidence intervals for these parameters are given by:
\begin{gather}
    CI(\substack{\mu_0\\T})=\substack{\tilde{\mu_0}\\\tilde{T}}\pm t_{\frac{\alpha}{2},n-2}SE(\substack{\mu_0\\T})
\end{gather}
where $t_{\frac{\alpha}{2},n-2}$ is the critical value of the  t-distribution for $\alpha$ confidence level and $n-2$ degrees of freedom and $n$ is the number of data points. The standard error $SE(\substack{\mu_0 \\ T}) = \sqrt{Cov(\mu_0, T)_{\substack{1,1 \\ 2,2}}} = \sqrt{(J^T J)^{-1} \sigma^2}$ is calculated from the covariance matrix. Here, $\sigma^2 = {\sum e_i^2}/{(n-2)}$ represents the variance, and $J$ is the Jacobian matrix defined by:
\begin{equation}
J = \begin{bmatrix}
\frac{\partial n(\mu_0,
T,x_1)}{\partial \mu_0} & \frac{\partial n(\mu_0,
T,x_1)}{\partial T} \\
\frac{\partial n(\mu_0,
T,x_2)}{\partial \mu_0} & \frac{\partial n(\mu_0,
T,x_2)}{\partial T} \\
... & ...
\\
\frac{\partial n(\mu_0,
T,x_n)}{\partial \mu_0} & \frac{\partial n(\mu_0,
T,x_n)}{\partial T} \\
\end{bmatrix}.
\end{equation}

From the measured density EOS, we are able to derive the local pressure P based on the Gibbs-Duhem relation $dP=nd\mu+SdT$. The pressure EoS is an integral of density with respect to the chemical potential \cite{PhysRevLett.119.165701},
\begin{equation}
    P(\mu,T)=\int_{\mu_c}^\mu n(\mu^\prime,T)d\mu^\prime.
\end{equation}
Here $\mu_c$ denotes the cut-off of this integration, which is restricted by the signal-to-noise ratio of our experiment measurement. Meanwhile, the measurement uncertainty of density distribution will propagate to the pressure, which gives 
\begin{equation}
\textrm{Var}[p(\mu,T)]=\sum_{\mu_i=\mu_c}^{\mu_i=\mu}\Delta\mu_i^2\textrm{Var}(n_i).
\end{equation}

\begin{figure*}[t]
\centering
\includegraphics[width=0.6\textwidth]{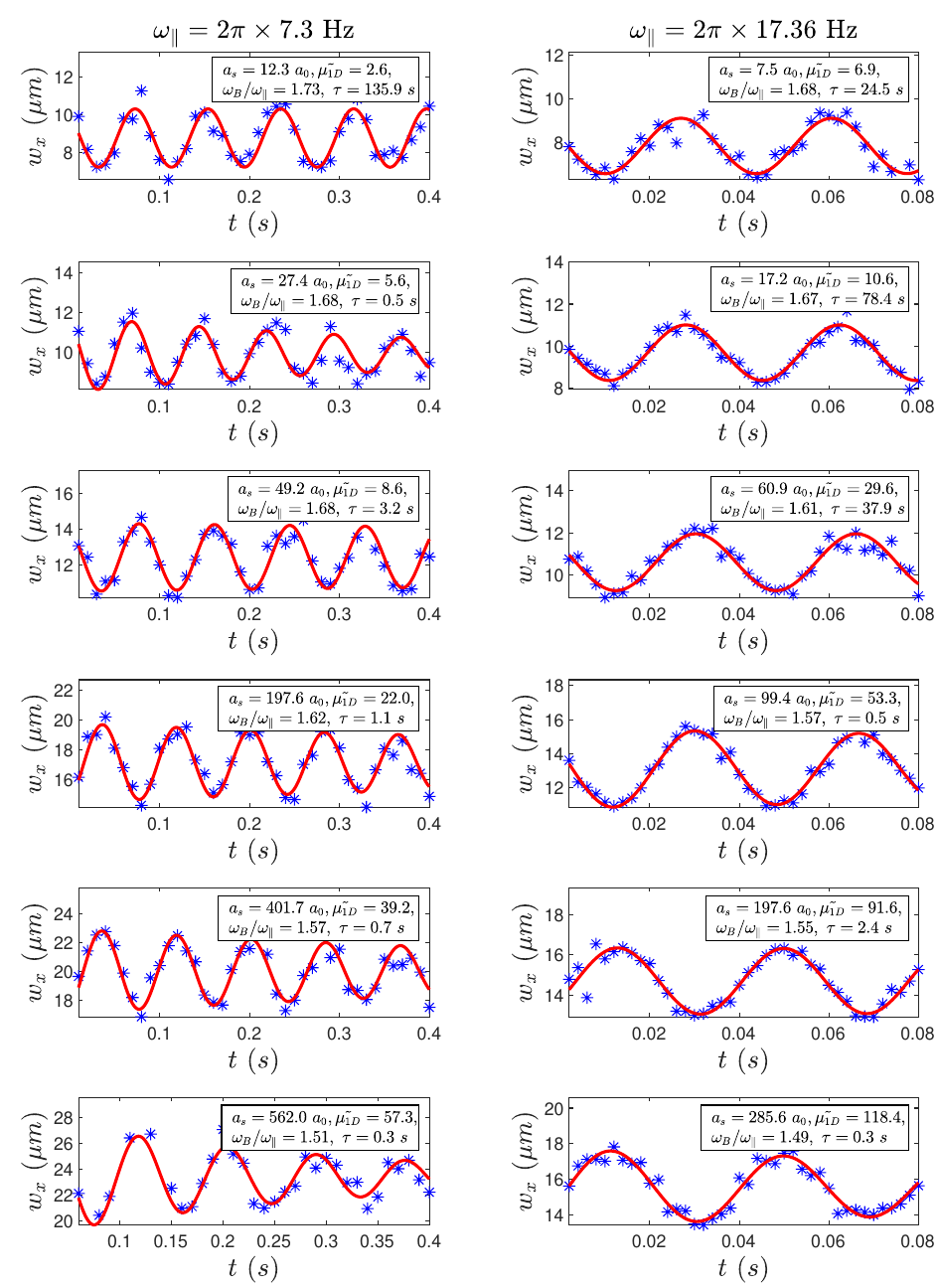}
\caption{Data for the measurements of breathing-mode frequencies. The blue stars correspond to measured $w_x$ and the red solid lines correspond to the fitting curves $w_x(t)=A\exp(-t/\tau) \sin(\omega_B t + \phi)+B$.
The left column shows data under $\omega_\parallel=2\pi\times$7.3~Hz, and the right column shows data under $\omega_\parallel=2\pi\times$17.36~Hz. 
From the top line to the bottom one, the scattering length $a_s$ increases. In each panel, we list the values of $a_s$, fitted $\omega_B/\omega_\parallel$, and $\tau$. 
}
\label{Fig.S1}
\end{figure*}

\section{Measurements of breathing-mode frequencies}
\label{Appendix:BreathingModeFreq}
We excite the breathing modes by quenching the scattering length $a_s$ \cite{PhysRevLett.77.988,PhysRevLett.77.420,PhysRevLett.113.035301}. Consequently, the width $w_x$ of the atomic cloud exhibits periodic oscillations. To analyze these oscillations, we fit the temporal evolution of the cloud's width $w_x$ using an exponentially decaying sinusoidal function, $w_x(t)=A\exp(-t/\tau) \sin(\omega_B t + \phi)+B$. This fitting procedure allows us to extract the frequency $\omega_B$ of the breathing mode and the damping time scale $\tau$ of the oscillations. In Fig.~\ref{Fig.S1}, we display representative data sets of the breathing modes under various scattering lengths $a_s$. For small scattering lengths, the oscillations closely resemble a pure sinusoidal form, indicating stable dynamics. Conversely, at larger scattering lengths, the oscillations exhibit a faster decay, suggestive of enhanced three-body losses affecting the system dynamics.

\begin{figure*}[t]
\centering
\includegraphics[width=1\textwidth]{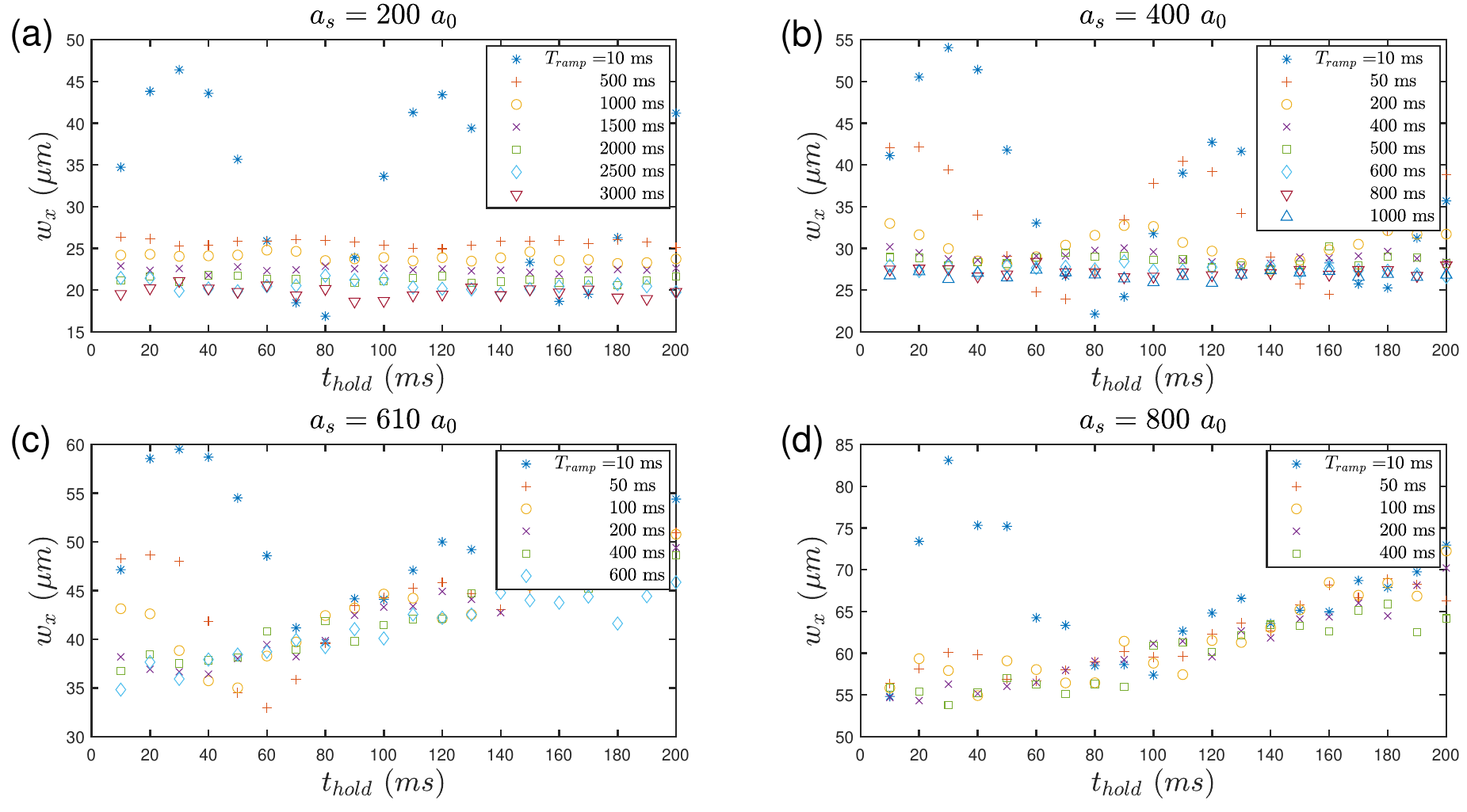}
\caption{The atomic cloud width $w_x$ versus the holding time $t_{hold}$ at different ramp time $T_{\text{ramp}}$ and scattering length $a_s$. Panel (a) to panel (d) correspond to $a_s=200a_0$, $400a_0$, $610a_0$, and 800$a_0$ respectively. In each panel, we list data for different $T_{ramp}$. For a very short ramp time such as $T_{ramp}$=10~ms, $w_x$ oscillates with the time $t_{hold}$ and it suggests that the ramp is too fast to satisfy the adiabatic theorem. When $T_{ramp}$ increases, the oscillation is suppressed. We list the oscillation amplitudes versus $T_{ramp}$ in Fig.~\ref{Fig.S3}. 
For a large scattering length such as 610$a_0$ or 800$a_0$, $w_x$ increases with respect to $t_{hold}$ and this shows the heating effects due to the three-body loss.
}
\label{Fig.S2}
\end{figure*}

\section{Three-body loss and associated heating}
\label{Appendix:3BodyLoss}
Three-body loss, especially prominent at large scattering lengths, substantially introduces heating, and constrains the achievable parameters in our experiments. 
After producing the BEC, we adiabatically transfer the atoms into the final 1560-nm elongated trap at a designated scattering length. The requirement of adiabaticity will set a minimum ramp time, which can be confirmed by the absence of breathing-mode excitations following the ramp procedure. 
On the other hand, for a very long ramp time, the heating effects due to three-body loss also become significant particularly at large scattering lengths $a_s$, and this also sets a time upper limit for our ramp procedure. 
Therefore, we need to balance these contributions to decide the ramp time $T_{ramp}$.
We evaluate these effects by measuring the condensate fraction and the atomic cloud width after the ramp, where the condensate fraction corresponds to how many atoms are in the ground state of radial directions according to the Yang-Yang equation. If the condensate fraction remains below 80\% even with the optimized ramp time, we consider that the corresponding scattering length cannot be achieved in our setup.

\begin{figure}[t]
\centering
\includegraphics[width=0.5\textwidth]{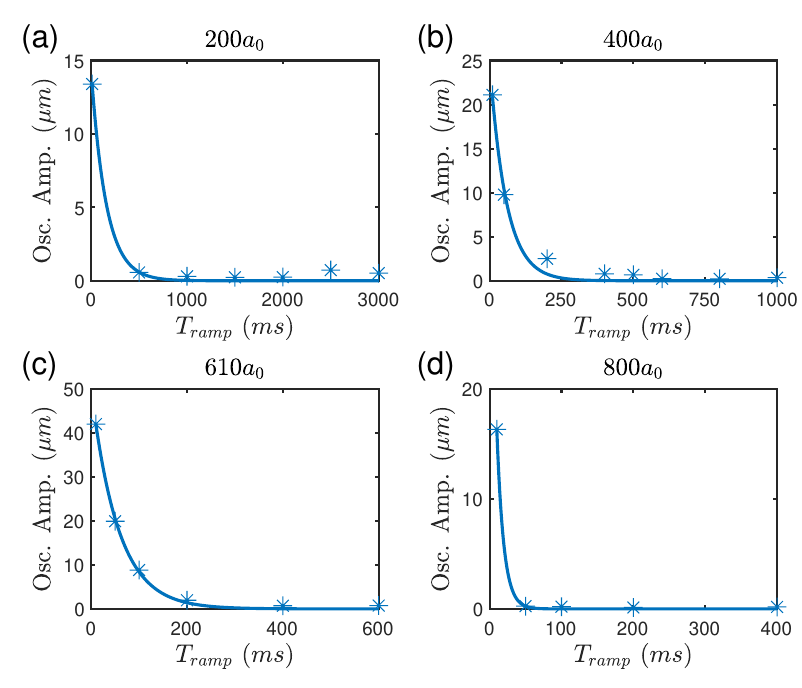}
\caption{Oscillation amplitudes versus different $T_{ramp}$ and scattering length $a_s$. From panel (a) to (d), they correspond to different scattering length $a_s=200a_0$, $400a_0$, $610a_0$, and 800$a_0$.
Then we choose the optimized ramp time at 1000, 600, 400, and 100~ms for 200$a_0$, 400$a_0$, 610$a_0$, and 800$a_0$ respectively, and the corresponded 1D density distributions are shown in Fig.~\ref{Fig.S4}.
The blue solid lines are fitting with an exponential decay function.
}
\label{Fig.S3}
\end{figure}
\begin{figure}[h]
\centering
\includegraphics[width=0.5\textwidth]{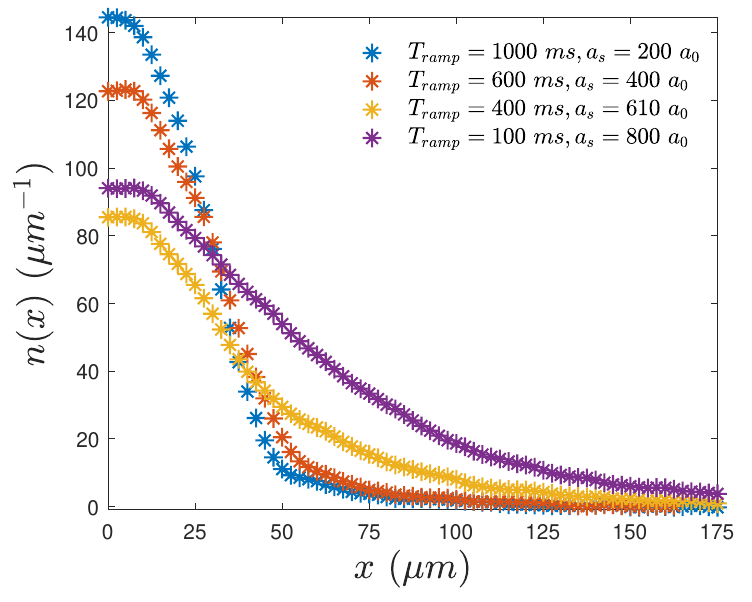}
\caption{The 1D density distributions at optimized $T_{ramp}$ and different scattering length $a_s$.}
\label{Fig.S4}
\end{figure}

\begin{figure}[hbtp]
\centering
\includegraphics[width=0.5\textwidth]{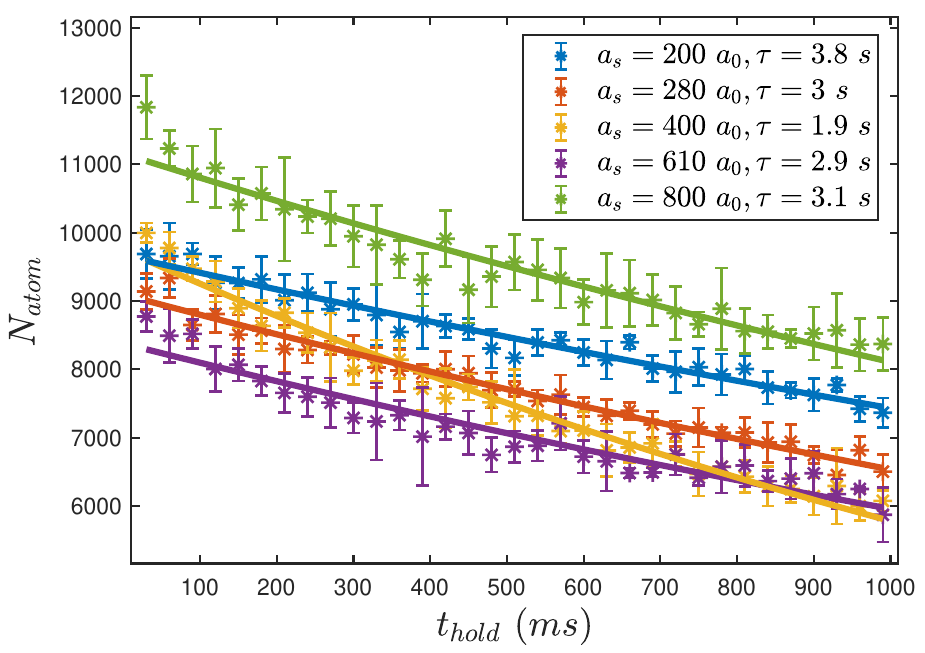}
\caption{The atom loss after the ramp procedure. The solid line represents an exponential-decay fit using $A \cdot \exp(-t/\tau)$.}
\label{Fig.S5}
\end{figure}

For example, at $\omega_\parallel=2\pi\times 7.3$~Hz and a given scattering length $a_s$, we utilize different ramp time $T_{ramp}$ to transfer atoms, and then hold atoms in the elongated dipole trap for a holding time $t_{hold}$. Then we measure the atomic cloud width $w_x$ for different combinations of $T_{ramp}$ and $t_{hold}$ at each scattering length $a_s$. 
The corresponding data is shown in Fig.~\ref{Fig.S2}. We can find that when $T_{ramp}$ is very small such as 10~ms, the atomic cloud width $w_x$ will oscillate versus the holding time $t_{hold}$. 
This suggests that the the time is too fast and the adiabatic theorem fails while the breathing mode is excited.
To more quantitatively analyze this, we fit this oscillation using a sinusoidal function to extract out the oscillation amplitudes which are shown in Fig.~\ref{Fig.S3}. 
In Fig.~\ref{Fig.S3}, we can see that the oscillation amplitudes quickly decay when $T_{ramp}$ increases.
Therefore, we can identify an optimized ramp time $T_{ramp}$ which is short for the three-body loss and also long enough to reduce the oscillation. 
At this setting, we analyze the density distributions and determine the condensate fraction by fitting it with the modified Yang-Yang equation (Fig.~\ref{Fig.S4}). If the resultant fraction is low ($<80\%$), this indicates excessive three-body heating, which compromise the feasibility of experiments at such large scattering lengths. Consequently, this restricts our achievable scattering lengths to $a_s < 285a_0$ for $\omega_\parallel=2\pi\times 3$~Hz and $a_s < 400a_0$ for $\omega_\parallel=2\pi\times 7.3$~Hz and $2\pi\times 17.36$~Hz. The atom loss after the ramp procedure is depicted in Fig.~\ref{Fig.S5}. Because we use different experimental conditions for different scattering lengths, the typical decay time based on an exponential function is not monotonic with the scattering length. However, it still provides a typical timescale to characterize the system with a large scattering length ($a_s\ge200 a_0$).

On the other hand, we can also inspect the heating effects due to three-body loss from Fig.~\ref{Fig.S2}. For a large scattering length such as 610$a_0$ and 800$a_0$, the atomic cloud width $w_x$ is increasing when we increase the holding time $t_{hold}$ with a fixed ramp time $T_{ramp}$.
Particularly for a very small $T_{ramp}$, the width $w_x$ is trying to oscillate one or two cycles and then turning to an increasing trend.
This is saying that once we prepare the atomic sample, the sample itself will be heated up and thermalized due to the three-body loss. 
Meanwhile, the large scattering length will induce damping into the breathing-mode oscillation.
For comparison, this effect is not so obvious for a smaller scattering length such as 200$a_0$, while the oscillation can last for a longer time. These behaviors give consistent parameters' constraints for our system that we cannot tune the scattering length too large, otherwise the sample will be heated up.

We also calculate the collisional rate of each atom to ensure that the system maintains thermal equilibrium throughout the atom loss process.
According to the one-dimensional scattering theory \cite{PhysRevLett.81.938,PhysRevLett.103.150601}, the scattering amplitude of a one-dimensional boson gas is given by
\begin{equation}
    \begin{aligned}&f_{\mathrm{even}}(k_{z})= - \frac{1}{1 + ik_{z}a_{1\mathrm{D}} - \underbrace{(ik_{z}a_{\perp}/2) \bar{\mathcal{L}} (-k_{z}^{2}a_{\perp}^{2}/4)}_{\mathcal{O}\left((k_{z}a_{\perp})^{3}\right)}} ,\end{aligned}
\end{equation}
where 
\begin{equation}
    \bar{\mathcal{L}}\left(\epsilon\right)=\sum_{n=1}^\infty(-1)^n\frac{\zeta[(1+2n)/2](2n-1)!! \epsilon^n}{2^nn!} .
\end{equation}
And the reflection probability for one boson is $\mathcal{R} = 1-|1 + f_{\mathrm{even}}|^{2}$. Thus the collisional rate for each particle is $R_c=\bar{v}n_{1D}\mathcal{T}$ where $\bar{v}$ is the typical velocity of particles. For the case of the most significant atom loss shown in Fig. 3, where $a_s = 280a_0$, the typical parameters are: $n = 80/\mu$m and $\omega_\perp = 2\pi \times 230$ Hz. Using the Lieb-Liniger (LL) equation for a zero-temperature 1D boson gas, we calculate the mean momentum as $k_{\text{mean}} = 1.8 \times 10^6 \, \text{m}^{-1}$, corresponding to a kinetic energy of 9.25 nK, which qualitatively agrees with the typical energy scale shown in Fig.~\ref{Fig3}.
Then we can get $\mathcal{R}=2.5\times10^{-4}$ and $R_c=0.027/$ms. From Fig.~\ref{Fig.S5}, the typical decay time for the atom number decaying to $1/e$ is around 3 s. Therefore, in the case of $a_s = 280a_0$ in Fig.~\ref{Fig3}, each atom will collide approximately 80 times within the typical decay time. This exceeds the generally accepted threshold of 10 collisions required for thermal equilibrium and implies that, despite the atom loss, the system reaches sufficient thermal equilibrium.

\end{document}